\documentclass[aps, prl, superscriptaddress, reprint]{revtex4-1}

\usepackage{siunitx}
\usepackage{chemformula}
\usepackage{amsmath}
\usepackage{amssymb}
\usepackage{graphicx}
\usepackage{hyperref}
\usepackage{verbatim}
\usepackage{lipsum}
\usepackage{comment}

\newcommand{\mr}[1]{\mathrm{#1}}

\newcommand{\cf}[0]{cf.~}

\newcommand{\fref}[1]{Fig.~\ref{fig:#1}}

\newcommand{\Cref}[1]{Chapter~\ref{chap:#1}}
\newcommand{\cref}[1]{Ch.~\ref{chap:#1}}

\renewcommand{\eth}[0]{Department of Materials, ETH Zürich, 8093 Zürich, Switzerland}

\newcommand{\tud}[0]{Institut f{\"u}r Festk{\"o}rper- und Materialphysik, Technische Universit{\"a}t Dresden and Würzburg-Dresden Cluster of Excellence ct.qmat, 01062 Dresden, Germany}

\newcommand{\wmi}[0]{Walther-Mei{\ss}ner-Institut, Bayerische Akademie der Wissenschaften, 85748 Garching, Germany}
\newcommand{\tum}[0]{Physik-Department, Technische Universit{\"a}t M{\"u}nchen, 85748 Garching, Germany}
\newcommand{\tumexc}[0]{Munich Center for Quantum Science and Technology (MCQST), 80799 M{\"u}nchen, Germany}
\newcommand{\ukn}[0]{Department of Physics, University of Konstanz, 78457 Konstanz, Germany}
\newcommand{\tou}[0]{Institute for Materials Research, Tohoku University, Sendai 980-8577, Japan}
\newcommand{\touwpi}[0]{AIMR and CSRN, Tohoku University, Sendai 980-8577, Japan}
\newcommand{\zernike}[0]{Zernike Institute for Advanced Materials, Groningen University, Groningen, The Netherlands}
\newcommand{\fudan}[0]{State Key Laboratory of Surface Physics and  Institute for Nanoelectronic Devices and Quantum Computing, Fudan University, Shanghai 200433, China}
\newcommand{\zhangjiang}[0]{Zhangjiang Fudan International Innovation Center, Fudan University, Shanghai 201210, China}

\begin{document}

\title{Magnetization dynamics affected by phonon pumping }

\author{Richard Schlitz}
\email{richard.schlitz@mat.ethz.ch}
\affiliation{\eth}
\affiliation{\tud}
\author{Luise Siegl}
\affiliation{\ukn}
\affiliation{\tud}
\author{Takuma Sato}
\affiliation{\tou}
\author{Weichao Yu}
\affiliation{\fudan}
\affiliation{\zhangjiang}
\affiliation{\tou}
\author{Gerrit E. W. Bauer}
\affiliation{\tou}
\affiliation{\touwpi}
\affiliation{\zernike}
\author{Hans Huebl}
\affiliation{\wmi}
\affiliation{\tum}
\affiliation{\tumexc}
\author{Sebastian T. B. Goennenwein}
\affiliation{\ukn}
\affiliation{\tud}

\date{\today}

\begin{abstract}
  
\textquotedblleft Pumping" of phonons by a dynamic magnetization promises to extend the range and functionality of magnonic devices.
We explore the impact of phonon pumping on room-temperature ferromagnetic resonance (FMR) spectra of bilayers of thin yttrium iron garnet (YIG) films on thick gadolinium gallium garnet substrates over a wide frequency range. 
At low frequencies the Kittel mode hybridizes coherently with standing ultrasound waves of a bulk acoustic resonator to form magnon polarons that induce rapid oscillations of the magnetic susceptibility, as reported before.
At higher frequencies, the phonon resonances overlap, merging into a conventional FMR line, but with an increased line width. 
The frequency dependence of the increased line broadening follows the predictions from phonon pumping theory in the thick substrate limit.
In addition, we find substantial magnon-phonon coupling of a perpendicular standing spin wave (PSSW) mode. 
This evidences the importance of the mode overlap between the acoustic and magnetic modes, and provides a route towards engineering the magnetoelastic mode coupling.
\end{abstract}

\maketitle

Magnons and phonons are, respectively, the elementary excitations of the magnetic and atomic order in condensed matter. 
They are coupled by weak magnetoelastic and magnetorotational interactions, which can often simply be disregarded. 
However, recent experimental and theoretical research reveals that the magnon-phonon interaction may cause spectacular effects in (i) ferromagnets close to a structural phases transition such as Galfenol~\cite{Godejohann2020, Sato2021} or (ii) magnets with exceptionally high magnetic and acoustic quality such as yttrium iron garnet~\cite{Streib2018, An2020, Rueckriegel2020, Holanda2021, Rezende2021, Graf2021, An2021}. 

Magnons are promising carriers for future low-power information and communication technologies~\cite{Chumak2017a, Chumak2021}. 
The magnon-phonon interaction can benefit the functionality of magnonic devices by helping to control and enhance magnon propagation when coherently coupled into magnon polarons~\cite{An2020, An2021}.
On the other hand, magnon non-conserving magnon-phonon scattering is the main source of magnon dissipation at room temperature~\cite{Cherepanov1993a, Gilbert2004}. 

The study of magnon-phonon interactions in high-quality magnets has a long history~\cite{Pomerantz1961,Matthews1962, Kooi1963, Rowell1963,Kooi1964, Wigen1965, Kobayashi1973, Sunakawa1984}. 
The arrival of crystal growth techniques, strongly improved microwave technology, and discovery of new phenomena such as the spin Seebeck effect, led to a revival of the subject in the past few years, with emphasis on ultrathin films and heterostructures~\cite{Uchida2010, Bombeck2012, Kikkawa2016, Goryachev2019, Harii2019, Goryachev2020, An2020, Babu2020, Litvinenko2021, Holanda2021, Polulyakh2021}.

High-quality yttrium iron garnet (YIG) is an excellent material to study magnons and phonons. 
Thin films grow best on single-crystal substrates of gadolinium gallium garnet (GGG), a paramagnetic insulator that is magnetically inert at elevated temperatures. 
However, the acoustic parameters of GGG are almost identical to YIG such that phonons are not localized to the magnet and thus the substrate cannot be simply disregarded. 
Streib et al.~\cite{Streib2018} pointed out that magnetic energy can leak into the substrate by magnon-phonon coupling by a process called \textquotedblleft phonon pumping\textquotedblright\ and predicted that it should cause an increased magnetization damping with a characteristic non-monotonous dependence on frequency.

Phonon pumping has been experimentally observed in the  ferromagnetic resonance of YIG\ films on GGG substrates~\cite{An2020, Litvinenko2021, Polulyakh2021}. 
These experiments revealed coherent hybridization of the (uniform) Kittel magnon with standing sound waves extended over the whole sample. 
In YIG/GGG/YIG trilayers phonon exchange couples magnons dynamically over \SI{}{mm} distances~\cite{An2020,An2021}. 
However, the predicted increased damping due to phonon pumping and the coupling of other than the macro-spin Kittel magnon remained elusive.
The direct detection of the increased damping is challenging due to the presence of inhomogeneous FMR line broadening and the resulting changes of the resonance line shape, in particular in the low frequency regime for thin films~\cite{Dubs2020} or due to the presence of several modes in the resonance for thicker YIG films~\cite{Kittel1958, Yu1975, Schreiber1996, Klingler2014}.

In this Letter, we report FMR\ spectra of YIG/GGG bilayers over a large frequency range, demonstrating the coupling of magnons and phonons from the high cooperativity to the incoherent regime.
We reproduce the magnon polaron fine structure at low frequencies~\cite{An2020, Litvinenko2021}, and evidence the presence of the acoustic spin pumping effect on the magnetic dissipation predicted in Ref.~\cite{Streib2018} at higher frequencies.
The excellent agreement with an analytical model allows us to extract the parameters for the phonon pumping by the (even) Kittel mode in the strong coupling regime, and provides insights into the strong-weak coupling regime at higher frequencies. 
In addition, we observe that the magnon-phonon coupling strength also is characteristically modulated for an (odd) perpendicular standing spin wave mode.  This shows that the overlap integral between magnon and phonon modes governs the coupling strength, thus opening a pathway for controlling it.

\onecolumngrid

\begin{figure}[th]
    \includegraphics{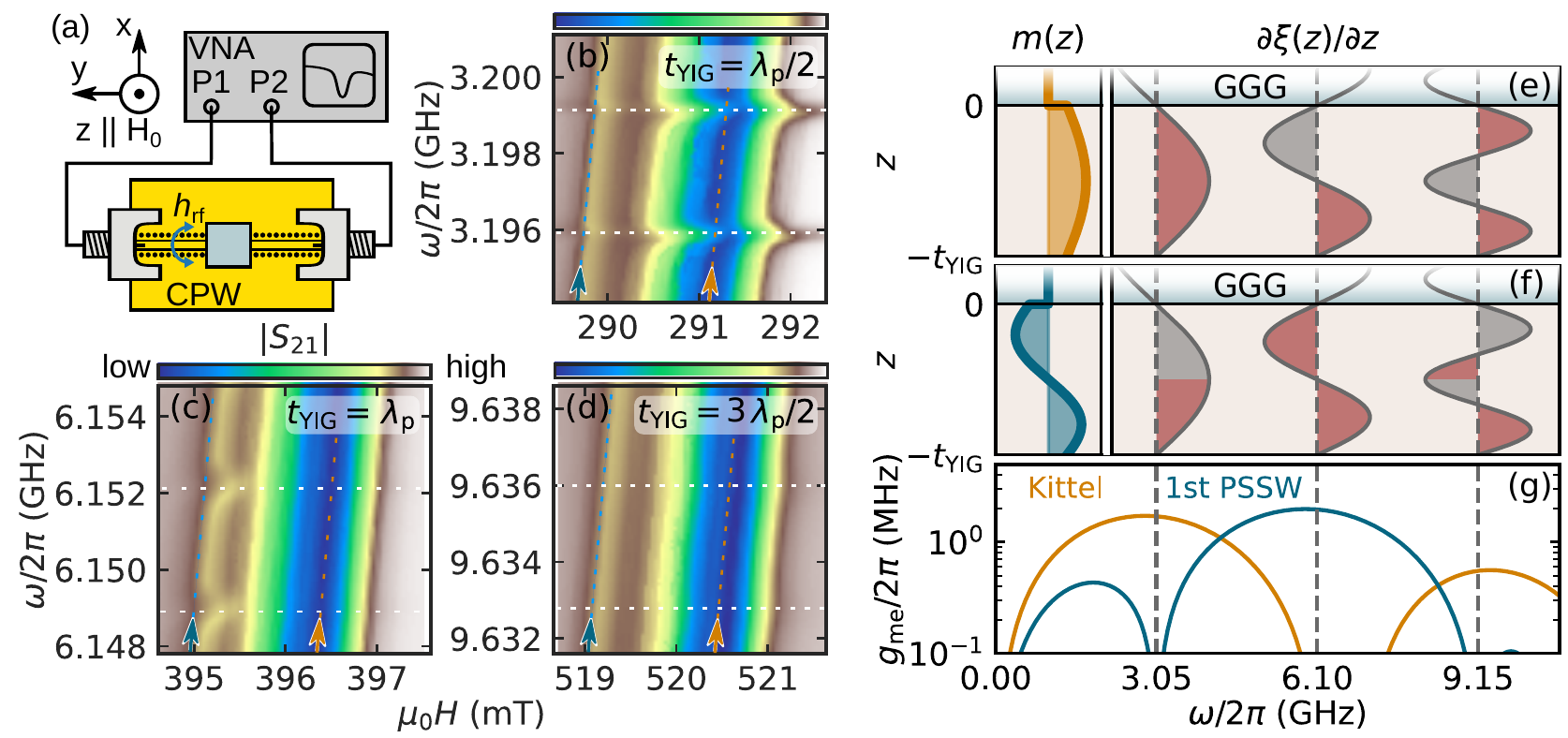}
    \caption{\label{fig:f1} 
    (a) A thin YIG film on a thick paramagnetic GGG substrate (gray square) is placed face down on a coplanar waveguide.
     The latter is connected to a vector network analyzer to obtain the transmission parameter $S_{21}$ as a function of frequency $\omega$.
     The external static magnetic field $H_0$ is applied normal to the surface.
        (b-d) High resolution maps of $|S_{21}|$ for different $\omega$ and $H_0$. 
        Shear waves with sound velocity $c_\mr{t}$ and wavelength $\lambda_\mr{p}$ form standing waves across the full layer stack.
        The three panels correspond to $t_\mr{YIG} \sim \lambda_\mr{p}/2, \lambda_\mr{p}$ and $3\lambda_\mr{p}/2$, respectively.
        The fundamental (Kittel) mode and the first perpendicular standing spin wave (PSSW) are marked with orange and blue dashed lines and arrows, respectively.
    (e,f) The thickness profile of the magnetic excitation $m_l(z)$ is shown for the Kittel mode (e) and the first PSSW (f) in the YIG film for $p = 0.5,$ together with the eigenmodes of the acoustic strain $\partial\xi(z)/\partial z$ corresponding to panels (b-d).
    The positive and negative contributions to the magnetoelastic exchange integral are shaded in red and gray, respectively.
    The magnetic excitation and thus the mode overlap vanishes in the GGG layer, whereas the phonons extend across full YIG/GGG sample stack.
    (g) The magnetoelastic mode coupling $g_{\mr{me}}$ is proportional to the overlap of the phonon and magnon modes and shows characteristic oscillations.
    }
\end{figure}
\twocolumngrid

Our sample consists of a \SI{630}{\nano\meter} \ch{Y3Fe5O12} film on a \SI{560}{\micro\meter} thick \ch{Gd3Ga5O12} substrate glued onto a coplanar waveguide (CPW) with a center conductor width $w=\SI{110}{\micro\meter}$.
It is inserted into the air gap of an electromagnet with surface normal parallel to the magnetic field [\cf \fref{f1}(a)]. 
We improve the magnetic field resolution to the \SI{1}{\micro\tesla} range by an additional Helmholtz coil pair in the pole gap of the electromagnet that is biased with a separate power supply.
We measure the complex microwave transmission spectra $S_{21}(\omega)$ by a vector network analyzer for a series of fixed magnetic field strengths over a large frequency interval at room temperature.

We first address $S_{21}(\omega)$ in the strong-coupling regime~\cite{An2020} in the form of  $|S_{21}|$ as a function of magnetic field and frequency, see \fref{f1}(b). 
The FMR reduces the transmission, emphasized by blue color and centered at the dashed orange line. 
Periodic perturbations in the FMR at fixed frequencies with period of $\sim\SI{3.2}{\mega\hertz}$ (dashed white lines) correspond to the acoustic free spectral range of the sample
\begin{equation}
\frac{\Delta\omega_{\mathrm{p}}}{2\pi}\approx\frac{c_{\mathrm{t}}}{2t_{\mathrm{GGG}}}\sim\SI{3.2}{\mega\hertz},\label{eq:ggg}%
\end{equation}
where $c_{\mathrm{t}}=\SI{3570}{\meter\per\second}$ is the transverse sound velocity of GGG~\cite{Ye1991}. 
These are the anticrossings of the FMR dispersion with field-independent standing acoustic shear wave modes across the full YIG/GGG layer stack~\cite{Comstock1963,Ye1988,An2020,Litvinenko2021, Polulyakh2021}. 
In \fref{f1}(b) we additionally observe a resonance corresponding to the PSSW (dashed blue line) shifted to a lower magnetic fields by exchange splitting $\mu_{0}\Delta H=D\pi^{2}/t_{\mathrm{YIG}}^{2}\sim\SI{1.4}{\milli\tesla}$, where $D=\SI{5e-17}{\tesla\meter^2}$ is the exchange stiffness of YIG~\cite{Kittel1958, Yu1975, Schreiber1996, Klingler2014}, but without visible coupling to the phonons. 

At $\SI{6.15}{\giga\hertz}$ [\cf\fref{f1}(c)] the periodic anticrossings vanish for the Kittel mode resonance, which implies a strongly suppressed magnon-phonon coupling. 
In contrast, the PSSW now exhibits clear anticrossings  similar to that of the Kittel mode in panel (b). 
Increasing the frequency further [\cf\fref{f1}(d)] to around \SI{9.63}{\giga\hertz}, the periodic oscillations in the PSSW vanish again, but the anticrossings of the Kittel mode do not recover. 

We interpret the suppression of the magnon polaron signal at higher frequencies around \SI{9}{\giga\hertz} in terms of a transition from the (underdamped) high cooperativity~\cite{An2020} to the (overdamped) weak coupling regime.
In the latter, the different phonon modes overlap, leading to a constant contribution of the phonon pumping to the magnon line width.
As a consequence, the periodic magnon polaron signatures vanish in favor of a slowly varying additional broadening of the FMR line that was predicted theoretically in the limit of thick GGG substrates~\cite{Streib2018,Sato2021}.

The coupling between the elastic and the magnetic subsystems in a confined magnet scales with the overlap integral of the phonon and magnon modes~\cite{Streib2018, Litvinenko2021}. 
The profile of a PSSW with index $l$ can be modelled by
\begin{align}
m_{l}(z) &  =p\sin\left(  [l+1]\pi\lbrack z+t_{\mathrm{YIG}}]/t_{\mathrm{YIG}%
}\right)  +\nonumber\\
&  (1-p)\cos\left(  l\pi\lbrack z+t_{\mathrm{YIG}}]/t_{\mathrm{YIG}}\right)  ,
\end{align}
where $z\in\left[  -t_{\mathrm{YIG}},0\right]$ and $0\leq p\leq1\ $interpolates between free ($p=0$) and pinned ($p=1$) surface dynamics. 
Assuming free elastic boundary conditions, a shear wave across the full layer stack with amplitude $\xi$ and frequency $\omega$ creates a strain profile (disregarding the standing wave formation and thus the finite free spectral range) in the YIG film that is given by

\begin{equation}
\frac{\partial\xi(z)}{\partial z}=\frac{\omega}{\widetilde{c_{\mathrm{t}}}%
}\sin\frac{\omega(t_{\mathrm{YIG}}+z)}{\widetilde{c_{\mathrm{t}}}},
\end{equation}
where $\partial\xi(z)$ is the local displacement and $\widetilde{c}_{\mathrm{t}}=\SI{3843}{\meter\per\second}$ is the transverse sound velocity of YIG. 
Note that the ladder of modes is disregarded here for simplicity.
The overlap integral of the fundamental (Kittel) mode with $l=0$ (Fig.~\ref{fig:f1}(e)) and the first PSSW with $l=1$ (Fig.~\ref{fig:f1}(f)) enters the interaction magnetoelastic coupling $g_{\mathrm{me}}$ as~\cite{Litvinenko2021}
\begin{equation}\label{eq:exchpssw}
g_{\mathrm{me},l}=\sqrt{\frac{2b^{2}\gamma}{\omega\rho M_{\mathrm{s}%
}t_{\mathrm{GGG}}t_{\mathrm{YIG}}}}\left\vert \int_{-t_{\mathrm{YIG}}}%
^{0}m_{l}(z)\frac{\partial\xi(z)}{\partial z}dz\right\vert ,
\end{equation}
where the parameters for YIG at room temperature are the magnetoelastic coupling constant $b=\SI{7e5}{\joule\per\meter^3}$, the mass density $\rho=\SI{5.1}{\gram\per\centi\meter^3}$, the gyromagnetic ratio $\gamma/2\pi=\SI{28.5}{\giga\hertz\per\tesla}$ and the saturation magnetization $M_\mathrm{s}=\SI{143}{\kilo\ampere\per\meter}$~\cite{An2020}.
$g_{\mathrm{me},0}$ and $g_{\mathrm{me},1}$ in Fig.~\ref{fig:f1}(g) for $p=0.5$ (yellow and blue lines, respectively) reveal differences in the magnetoelastic coupling of the different magnetic modes. 
In both cases the coupling oscillates as a function of frequency, but the maxima for $l=0$ and $l=1$ are shifted by $l\cdot\widetilde{c}_\mr{t}/2 t_\mr{YIG} \approx \SI{3}{\giga\hertz}$.
Note that this is true also for the higher standing spin wave modes, so that even at high frequencies, strong magnon-phonon interactions can be realized.
In other materials the results may depend on the details of the interface and surface boundary conditions~\cite{Bombeck2012}.

\begin{figure}[th]
\includegraphics{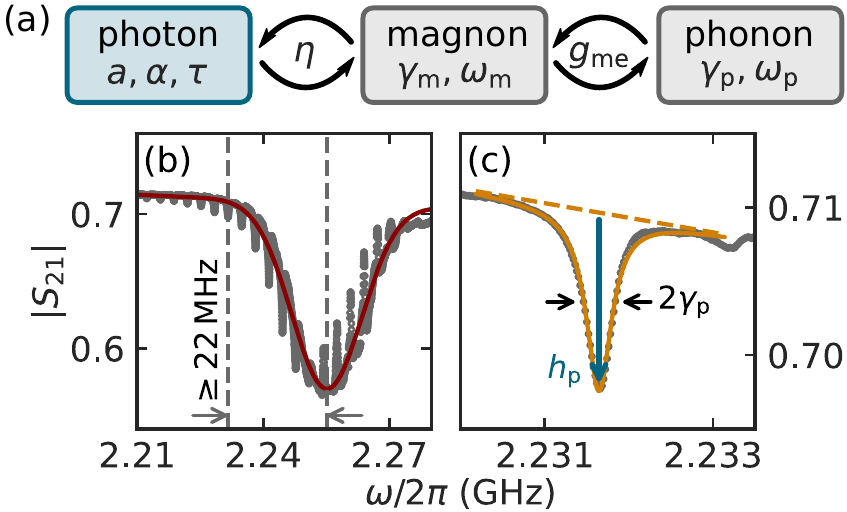} 
\caption{(a) The phonons and magnons in YIG/GGG bilayers form a two-partite system that can be modeled as coupled harmonic oscillators that are driven by microwave photons~\cite{An2020}.
The parameters are the resonance frequencies $\omega_{\mathrm{m}},\omega_{\mathrm{p}}$, damping constants $\gamma_{\mathrm{m}},\gamma_{\mathrm{p}},$ and coupling strength $g_{\mathrm{me}}$. 
The coplanar waveguide with  transmission amplitude $a$, phase $\alpha$ and electric length $\tau$, interacts with the magnon system parametrized by the coupling strength $\eta$. 
(b) $|S_{21}|$ as a function of the frequency for $\mu_{0}H\approx\SI{259}{\milli\tesla}$. A Gaussian fit (red line) determines the FMR frequency $\omega_{\mathrm{m}}$ (right dashed line). 
(c) Zoom-in on the phonon line with $\omega_{\mathrm{m}}-\omega_{\mathrm{p}} = \SI{22}{\mega\hertz}$ (marked by the left dashed line in panel a). 
We obtain the line width $\gamma_{\mathrm{p}}$ and the amplitude $h_{\mathrm{p}}$ of the phonon resonance by a Lorentzian fit. }%
\label{fig:f2}%
\end{figure}

A phonon and a magnon mode with discrete frequencies $\omega_{\mathrm{p}}$ and $\omega_{\mathrm{m}}$ $\left[  =\gamma\mu_{0}(H-M_{\mathrm{eff}})\text{ for the Kittel mode}\right]$ and amplitudes $A_{\mathrm{p}}$ and $A_{\mathrm{m}},$ respectively, behave as two harmonic oscillators coupled by the interaction $g_{\mathrm{me}}$~\cite{An2020}:
\begin{align}
-A_{\mathrm{m}}(\gamma_{\mathrm{m}}+i(\omega_{\mathrm{m}}-\omega
))-iA_{\mathrm{p}}g_{\mathrm{me}}/2+\eta &  =0\\
-A_{\mathrm{p}}(\gamma_{\mathrm{p}}+i(\omega_{\mathrm{p}}-\omega
))-iA_{\mathrm{m}}g_{\mathrm{me}}/2 &  =0,\label{eq:coupled}%
\end{align}
where $\gamma_{\mathrm{m}}$ and $\gamma_{\mathrm{p}}$ are the decay rates (in angular frequency units). 
$\eta$ parametrizes the coupling of the magnetic order to the external microwaves at frequency $\omega$, see Fig.~\ref{fig:f2}(a).
The solution for the magnetic amplitude is
\begin{equation}
A_{\mathrm{m}}=\eta\left[  \left(  \frac{g_{\mathrm{me}}}{2}\right)  ^{2}%
\frac{1}{\gamma_{\mathrm{p}}+i(\omega_{\mathrm{p}}-\omega)}+(\gamma
_{\mathrm{m}}+i(\omega_{\mathrm{m}}-\omega))\right]  ^{-1}.\label{eq:fit}%
\end{equation}
This resonator couples to a CPW according to~\cite{Probst2015}
\begin{equation}
S_{21}(\omega)=a\exp(i\alpha)\exp(-i\tau\omega)\left[1-A_{\mathrm{m}}\right].\label{eq:S21}
\end{equation}
in which the first part in the square brackets represents the external circuit with frequency-dependent amplitude and phase shift $a$ and $\alpha,$
respectively, and $\tau$ is an electronic delay time. 
We can fit the unknown parameters $\eta$ and $g_{\mathrm{me}}$ to the observed spectra in
Fig.~\ref{fig:f2}(b,c)]. 
$\eta$ can be extracted from the amplitude of the ferromagnetic resonance by solving $h_{\mathrm{m}}=\left\vert S_{21}(\omega=\omega_{\mathrm{m}})_{\eta=0}\right\vert -\left\vert S_{21}(\omega=\omega_{\mathrm{m}})_{g_{\mathrm{me}}=0}\right\vert \equiv f(\eta)$.
Similarly, the data for a phonon resonance  $h_{\mathrm{p}}=\left\vert S_{21}(\omega=\omega_{\mathrm{p}})_{g_{\mathrm{me}}=0}\right\vert -\left\vert S_{21}(\omega=\omega_{\mathrm{p}})\right\vert \equiv g(g_{\mathrm{me}})$ can be solved for $g_{\mathrm{me}}$.
$h_{\mathrm{p}}$ and $h_{\mathrm{m}}$ can be extracted from fits to the experimental data.

We fit the Kittel mode lines in $|S_{21}(\omega)|$ at different fixed magnetic fields by a Gaussian to distill the resonance frequency $\omega_{\mathrm{m}}$,
the amplitude $h_{\mathrm{m}}$ and width $\gamma_{\mathrm{m}}$ (cf.~Fig.~\ref{fig:f2}b). 
A good fit by a Gaussian line shape indicates inhomogeneous broadening of the FMR, see below. 
Next, we select an acoustic resonance at a frequency $\omega_{\mathrm{p,0}}$ with $(\omega_{\mathrm{m}}-\omega_{\mathrm{p,0}})/2\pi>2\gamma_{\mathrm{m}}/2\pi\approx\SI{22}{\mega\hertz}$, which is only weakly perturbed by the magnon-phonon coupling, but still has a significant oscillator strength.
For a better statistics, we independently fit a total of six phonon resonances with frequencies below $\omega_{\mathrm{p,0}}$  by Lorentzians [cf.~ Fig.~\ref{fig:f2}(c)] to extract their average height $h_{\mathrm{p}}$ and broadening $\gamma_{\mathrm{p}}$. 

\begin{figure}[th]
\includegraphics{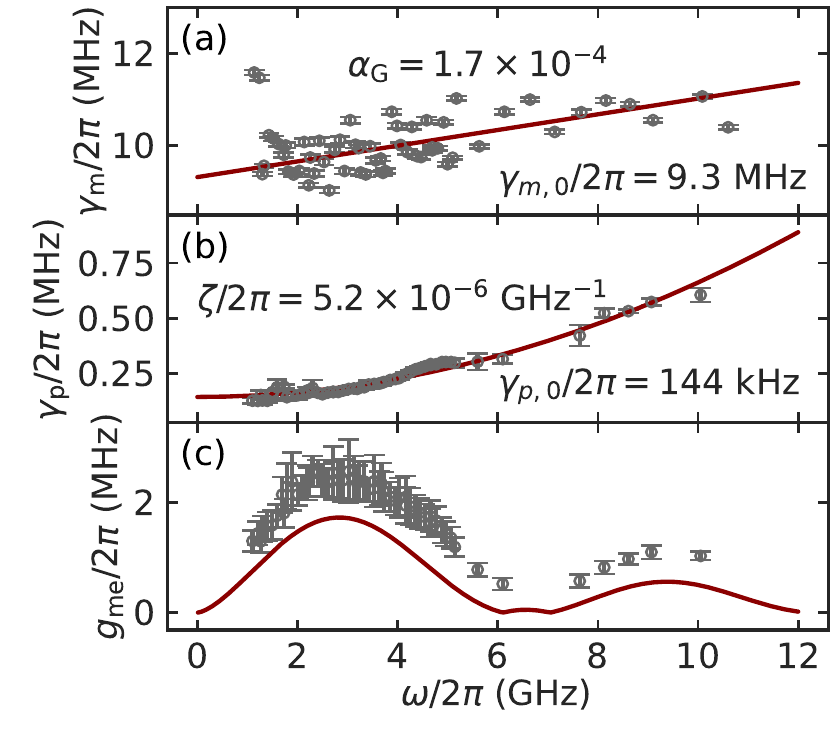} \caption{(a) Half width at half maximum (HWHM) obtained from Gaussian fits to the FMR lines. 
(b) HWHM from the Lorentzian fits to the acoustic resonances.
(c) Magnetoelastic mode coupling obtained from the harmonic oscillator model using the parameters from the fits shown in (a), (b) and in the SM~\cite{SMME}. 
The maximum coupling strength is $\sim\SI{2.2}{\mega\hertz}$. }%
\label{fig:f3}%
\end{figure}

The resulting fit parameters are summarized in Fig.~\ref{fig:f3} and in the supporting material~\cite{SMME}. 
The line width of the FMR $\gamma_{\mathrm{m}}=\gamma_{\mathrm{m,0}}+\alpha_{\mathrm{G}}\omega$ shown in panel (a) is dominated by inhomogeneous broadening $\gamma_{\mathrm{m,0}}/2\pi=\SI{9.3}{\mega\hertz}$, which we associate to variations of the local (effective) magnetization over the \SI{6}{\milli\meter} long sample and across the thickness profile. 
The low Gilbert damping $\alpha_{\mathrm{G}}\sim\SI{1.7e-4}{}$ is evidence for an intrinsically high quality of the YIG film. 
We associate the parabolic increase of the sound attenuation $\gamma_{\mathrm{p}}=\zeta \omega^{2}+\gamma_{\mathrm{p,0}}$ [cf.~Fig.~\ref{fig:f3}(b)] to thermal phonon scattering in GGG~\cite{Dutoit1972,Dutoit1974,Daly2009}.
The inhomogeneous phonon line width $\gamma_{\mathrm{p,0}}/2\pi=\SI{144}{\kilo\hertz}$ may be caused by a small angle $\sim\SI{1}{\degree}$ between the bottom and top surfaces of our sample~\cite{Krzesinska1984}, where the estimate is based on the phonon mean-free-path $\delta\sim c_{\mathrm{t}}/\gamma_{\mathrm{p}}\approx\SI{4}{\milli\meter}$~\cite{An2020}.
We do not observe a larger scale disorder in the substrate thickness that would contribute a term $\sim\omega$ to the attenuation~\cite{Dutoit1974}.

In the lower frequency regime $\omega/2\pi\lesssim\SI{10}{\giga\hertz}$ the phonon mean-free path $\delta>\SI{1}{\milli\meter}$ is larger than twice the thickness of the bilayer. 
At frequencies above $\SI{10}{\giga\hertz}$, however, we enter the cross-over regime between high cooperativity and weak coupling in which the phononic free spectral range approaches its attenuation ($\Delta\omega_\mathrm{p} \sim 2\gamma_\mathrm{p}$).
The fitting with individual phonon lines becomes increasingly inaccurate, as the baseline of the FMR signal without contributions due to phonons cannot be established from the data. 
In this regime, the strongly overlapping phonon lines thus lead to an average increase of the FMR line width in addition to the rapidly oscillating contributions. 
If the frequency is increased further, the mode overlap further increases and the thickness of the stack becomes irrelevant so that we can take it to be infinite.
In this limit, and for finite magnetoelastic coupling, phonons just give rise to an average broadening of the FMR line.
While indirect, our observations thus confirm the predicted damping enhancement by phonon pumping in the incoherent limit~\cite{Streib2018,Sato2021}.

The oscillations observed in magnetoelastic mode coupling $g_{\mathrm{me}}$ in panel \ref{fig:f3}(c) agree well with the model Eq.~(\ref{eq:exchpssw}) (red
line) for a YIG film with a thickness of $t_{\mathrm{YIG}}=\SI{630}{\nano\meter}$ and a pinning parameter $p=0.5$ (from Fig.~\ref{fig:f1}(g)) for $\omega/2\pi\lesssim\SI{7}{\giga\hertz}$.
An alternative assessment based on a full fit of the experiments by the coupled equations for the complex scattering parameter leads to a similar $g_{\mathrm{me}}/2\pi=\SI{1.6}{\mega\hertz}$ at $\omega/2\pi\approx\SI{2.2}{\giga\hertz}$ (see SM~\cite{SMME}).
The model likely overestimates the coupling strength, since the inhomogeneous contributions to the line broadening are not considered independently here.

In summary, our high resolution FMR data taken over a broad frequency range confirm that magnon-phonon coupling in confined systems depends not only on the material parameters, but also qualitatively changes with the mode overlap. 
This provides the option of tuning the magnon-phonon coupling strength by the frequency, magnetic field variations and sample geometry. 
We analyzed the magnon-phonon mode coupling over a broad frequency range by a simple harmonic oscillator model, revealing the  oscillating nature of the acoustic spin pumping efficiency as predicted theoretically~\cite{Streib2018}.
Broadband phonon pumping experiments in heterostructures as presented here can thus be used as experimental platform to study the influence of the magnetic phase diagram on the acoustic properties also in an adjacent \emph{magnetic} substrate, e.g.~in the frustrated magnetic phase at very low temperatures in GGG~\cite{Deen2015}.

\acknowledgments
We would like to thank O.~Klein and A.~Kamra for fruitful discussions. We acknowledge financial support by the Deutsche Forschungsgemeinschaft via SFB 1432 (project no.~B06), SFB 1143 (project no.~C08), the Würzburg-Dresden Cluster of Excellence on Complexity and Topology in Quantum Matter - ct.qmat (EXC 2147, project-id 39085490), and the Cluster of Excellence ``Munich Center for Quantum Science and Technology'' (EXC 2111, project-id 390814868).

\bibliography{190618_bibliography.bib}

\end{document}